\documentstyle[jkas35]{article}

\runningauthor{X. ZHANG}
\runningtitle{SECULAR EVOLUTION OF GALAXIES}

\beginpage{1}
\endpage{??}

\begin{document}

\title{SECULAR EVOLUTION OF SPIRAL GALAXIES}

\author{Xiaolei Zhang}

\address{US Naval Research Laboratory, 
Remote Sensing Division, \\
4555 Overlook Ave., SW, Washington, DC 20375, USA \\
{\it E-mail: xiaolei.zhang@nrl.navy.mil}}


\abstract{
It is now a well established fact that galaxies undergo significant
morphological transformation during their lifetimes, manifesting as an
evolution along the Hubble sequence from the late to the early Hubble
types.  The physical processes commonly believed to be responsible for
this observed evolution trend, i.e. the major and minor mergers, 
as well as gas accretion under a barred potential, though demonstrated 
applicability to selected types of galaxies, on the whole have failed to 
reproduce the most important statistical and internal properties of
galaxies.  The secular evolution mechanism reviewed in this paper 
has the potential to overcome most of the known difficulties of the 
existing theories to provide a natural and coherent explanation of
the properties of present day as well as high-redshift galaxies.}

\maketitle

\section{INTRODUCTION}

It is a generally accepted view that our understanding of 
the mechanisms and processes responsible for the formation and 
evolution of galaxies is incomplete.  However, the degree of 
this incompleteness does not come into sharp focus until we 
assemble our most comprehensive and up-to-date observational 
knowledge, both on the large-scale distribution 
as well as on the internal properties of galaxies, and 
compare these with the predictions of our working theoretical 
models to try to make a coherent picture. 

Take our own home galaxy, the Milky Way, as an example.  It is a
typical field galaxy of Sbc type (de Vaucouleurs \& Pence 1978)
in a small group environment.  Existing theories offer several possible 
ways for the formation of this type of galaxy.  The scenario
offered by the earliest monolithic collapse model 
(Eggen, Lynden-Bell, \& Sandage 1962) is
that the Galaxy's mass distribution acquired most of its shape
from the very beginning, i.e. about a Hubble time ago, and 
subsequently underwent only passive luminosity evolution driven
by star formation, nucleosynthesis and element recycling.  One 
problem with this static picture is that the observed kinematics of the 
different age groups of stars in the Milky Way disk differ 
systematically, manifesting as the well-known age-velocity dispersion 
relation of the solar neighborhood stars (Wielen 1977).  In the 
primordial collapse model, there is nothing which could account 
for this secular increase of the velocity dispersion of disk stars.  

Recent deep surveys have found that galaxies in the general field
environment similar to that occupied by the Milky Way have 
undergone significant morphological transformation over the cosmic
time, following a similar trend though not to as dramatic a degree 
as the cluster galaxies we will discuss next.  It is found that 
more field galaxies are of earlier Hubble types in the nearby universe 
than at the higher redshifts (Lilly et al. 1998).  There is also a 
population of so-called faint-blue galaxies, which are in fact $L_*$ 
galaxies having luminosities and sizes similar to the Milky Way, 
which exist at the intermediate redshifts but which have all but 
disappeared in the nearby universe (Ellis 1997).

Could this observed morphological evolution be due to the hypothesized
major/minor merger events?  The thinness of the Milky Way disk, 
the lack of a large population of counter rotating stars, 
as well as the smoothness of the age-velocity dispersion relation 
(Figure 1) all argue against either a major merger or 
the accretion of a satellite of significant mass over the past Hubble 
time (Ostriker 1990; Wyse 2001).  There is indeed a known discontinuity 
in the Galaxy age-velocity dispersion relation at about 11 Gyr ago
(Binney, Dehnen, \& Bertelli 2000; Gilmore, Wyse, \& Norris 2002).  
Although a merger has been proposed as its origin, 
the emergence of the spiral structure on the disk at about the
same time seems a more likely cause for this discontinuity and
for the creation of the thick disk.  Furthermore, the Milky Way bulge
stellar populations are distinctively different from its 
known satellites such as the Magellanic clouds (Gilmore 2001),
further arguing against the building of a significant fraction of the
Bulge through satellite accretion.

\begin{figure}[!htbp]
\begin{center}
\epsfysize=2.3in
\epsffile{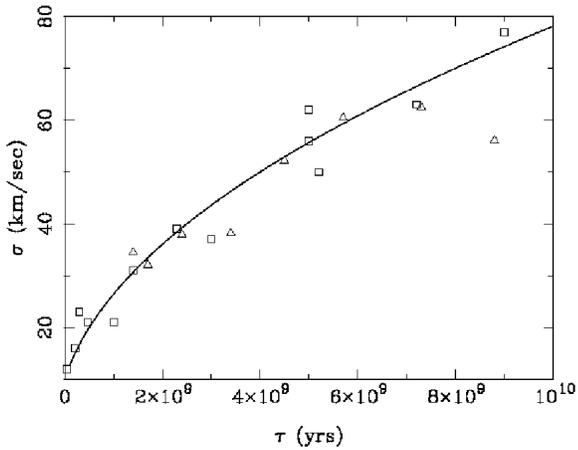}
\bigskip
\caption{The age-velocity dispersion relation of the solar neighborhood
stars. Points: observational data [squares: Wielen (1977); triangles:
Carlberg et al. (1985)]. Curve: fitted disk star heating 
due to spiral structure (Zhang 1999).} 
\end{center}
\end{figure}

Is the bar-driven gas accretion process (Kormendy 1982, 1993; 
Pfenniger \& Norman 1990; Combes et al. 1990; Pfenniger \& Friedli 1991;
Friedli \& Benz 1993; Norman, Sellwood \& Hasan 1996; Courteau,
de Jong \& Broeils 1996) responsible for Bulge building?  It was indeed
shown in numerical simulations that gas accretes towards the center after 
being shocked crossing the galactic bar potential; 
however, the abundance analyses of 
stars in the thick disk and bulge of the Milky Way indicate a high 
[$\alpha$/Fe] ratio (Gilmore 2001), where $\alpha$ elements O and Mg are 
created primarily in type II supernovae from massive young stars and the Fe is 
generated in type I supernovae from lower-mass stars of much greater ages.
The high [$\alpha$/Fe] abundance ratio can arise if the Bulge and
thick-disk stars all formed within a short time, thereby suppressing
enrichment in Fe which requires longer timescales.  These
timescale considerations suggest that most of the Bulge stars have formed
very early on and probably not far from their present locations.
This limits the importance of gaseous inflow in building the Bulge,
as continuous inflows would extend the star-forming
epoch and so enable Fe enrichment from the type I supernovae
(B. Waller 2001, private communication;
Jablonka, Gorgas \& Goudfrooij 2002).

We now turn our attention to dense clusters. This is in fact the
environment where the morphological transformation of galaxies was
first indicated through the so-called Butcher-Oemler (BO) effect 
(Butcher \& Oemler 1978a,b).  When it was discovered,
the BO effect referred to a bluing of colors for galaxies in the
dense clusters at the intermediate redshifts compared to similar
density clusters in the local universe, which contain mostly red
early type galaxies. Recent HST observations (Couch et al. 1994; 
Dressler et al. 1994) have been able to resolve the morphology 
of the BO galaxies ans show that they are mostly late type disks; 
therefore the BO effect is now considered not only
a color evolution effect but also a morphological transformation effect.
Major mergers are not likely to be responsible for the
observed morphological transformation of the Butcher-Oemler 
(BO) cluster galaxies, due to the high-speed nature of the 
encounters (Dressler et al. 1997), as well as the finding 
through numerical simulations that dissipationless mergers between 
preexisting stellar disks cannot account for the kinematics 
of the early type galaxies, 
especially the large ratio of the rotational to random velocities 
(Heyl, Hernquist, \& Spergel 1996; Cretton et al. 2001); whereas the 
spiral disks in BO clusters are found to have formed most of 
their stars before the observed morphological transformation 
had taken place (Franx \& van Dokkum 2001).  Minor mergers are also unlikely
to be the main driver because there does not seem to be 
a large reservoir of dwarf spheroidal satellites in these 
clusters (Trentham 1997) to cause the simultaneous morphological 
transformation of the large number of BO galaxies.

The rapid morphological changes of galaxies in clusters
are also not likely to be produced by a ram pressure gas-stripping
mechanism alone (Gunn \& Gott 1972), since stripping could
not lead to a change of bulge-to-disk ratio (Sandage 1983), 
whereas the morphological transformation of the BO galaxies
from the late type disks to S0s and ellipticals requires
such a change.  The so-called ``harassment'' mechanism had been 
shown to be effective in stripping away the outer gas and transforming 
the small late-type disks into early type dwarf galaxies (Moore et al. 1996),
yet it was shown to be much less effective on large disks (Gnedin 1999).

Thus we see, for both the field and cluster galaxies, the existing
theories are unable to provide satisfactory explanations of their
formation and evolution.

\section{SECULAR MORPHOLOGICAL EVOLUTION OF GALAXIES}

During the past few years, a new mechanism for the secular
evolution of galaxies has been proposed (Zhang 1996, 1998, 1999)
which operates through large scale coherent patterns in galaxies
such as spirals, bars or other skewed mass distributions.
This mechanism has been overlooked by past workers in this
field due to some subtle features of its operation.

\subsection{The Source and Sink of Outward Angular Momentum Transport}

Three decades have past since the publication of
Lynden-Bell \& Kalnajs' seminal paper (Lynden-Bell \&
Kalnajs 1972, hereafter LBK) demonstrating that spiral density
waves in disk galaxies can transport angular momentum (as well
as energy) outward.
Associated with this outward angular momentum transport
is an expected secular redistribution of disk matter,
coinciding with the trend of the entropy evolution of a
self-gravitating system; i.e., towards a more and
more centrally concentrated core together with
the build-up of an extended outer envelope.
However, this latter aspect has rarely been discussed, if at
all, in the context of the LBK theory since the publication
of their paper.  One of the reasons for
this disparity is the equally well-known second result from the
same paper: there is no interaction between a steady amplitude
spiral density wave and the basic state (i.e. the axisymmetric
part) of the galactic disk except at
the inner and outer Lindblad resonances (ILR and OLR).
There is thus an apparent lack of consistency in the LBK theory since
it is difficult to imagine a spiral wave constantly transporting
angular momentum outward without the basic state mass distribution
undergoing a corresponding change.  One possible way out
of the apparent contradiction is if the spiral pattern is a
transient phenomenon as was assumed by LBK.  
The outward angular momentum transport
then leads to a temporary and short-lived growth
of a wave train between the ILR and OLR: since the wave has
negative angular momentum density inside corotation
relative to the axisymmetric disk, the outward angular momentum
transport leads to its own spontaneous growth.

However, studies of grand design galaxies in groups
(Elmegreen \& Elmegreen 1983, 1989) suggest that most spiral patterns
last for at least 10 revolutions if they are triggered by
interactions.  Recent N-body simulations
have also produced long-lived spiral patterns in isolated stellar disks
which lasted more than 10 revolutions with essentially constant
pattern speed and wave amplitude (Figure 2).

\begin{figure*}[!htbp]
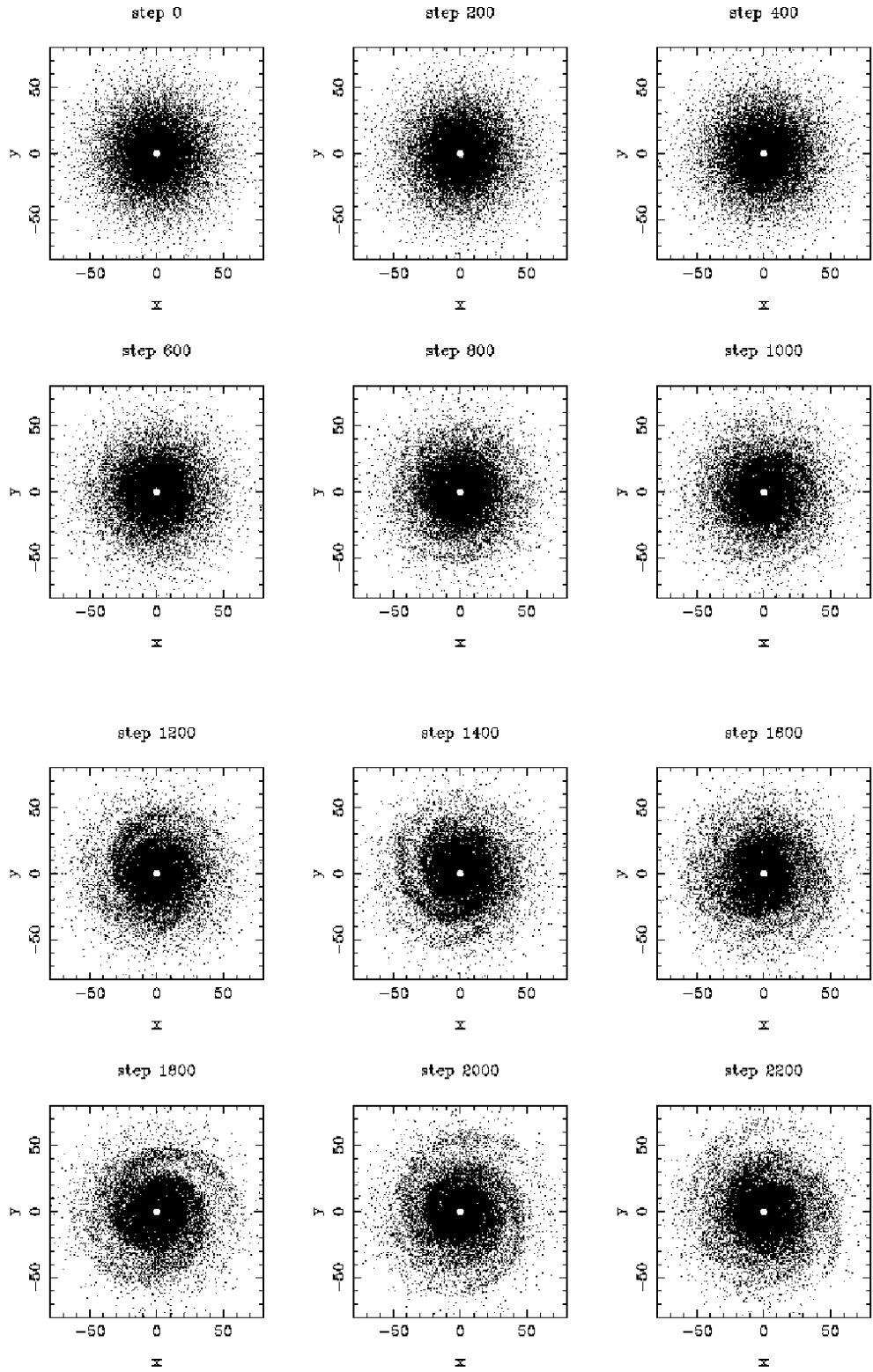

\begin{center}
\epsfysize=3.8in
\epsffile{figure1a.epsi}
\end{center}
\smallskip
\begin{center}
\epsfysize=3.8in
\epsffile{figure1b.epsi}
\bigskip
\caption{Morphological evolution of an N-body spiral mode
in a purely stellar disk.  Between the adjacent frames
the pattern rotates about $120^o$.  From Zhang (1998).}
\end{center}
\end{figure*}

\setcounter{figure}{1}
\begin{figure*}[!htbp]
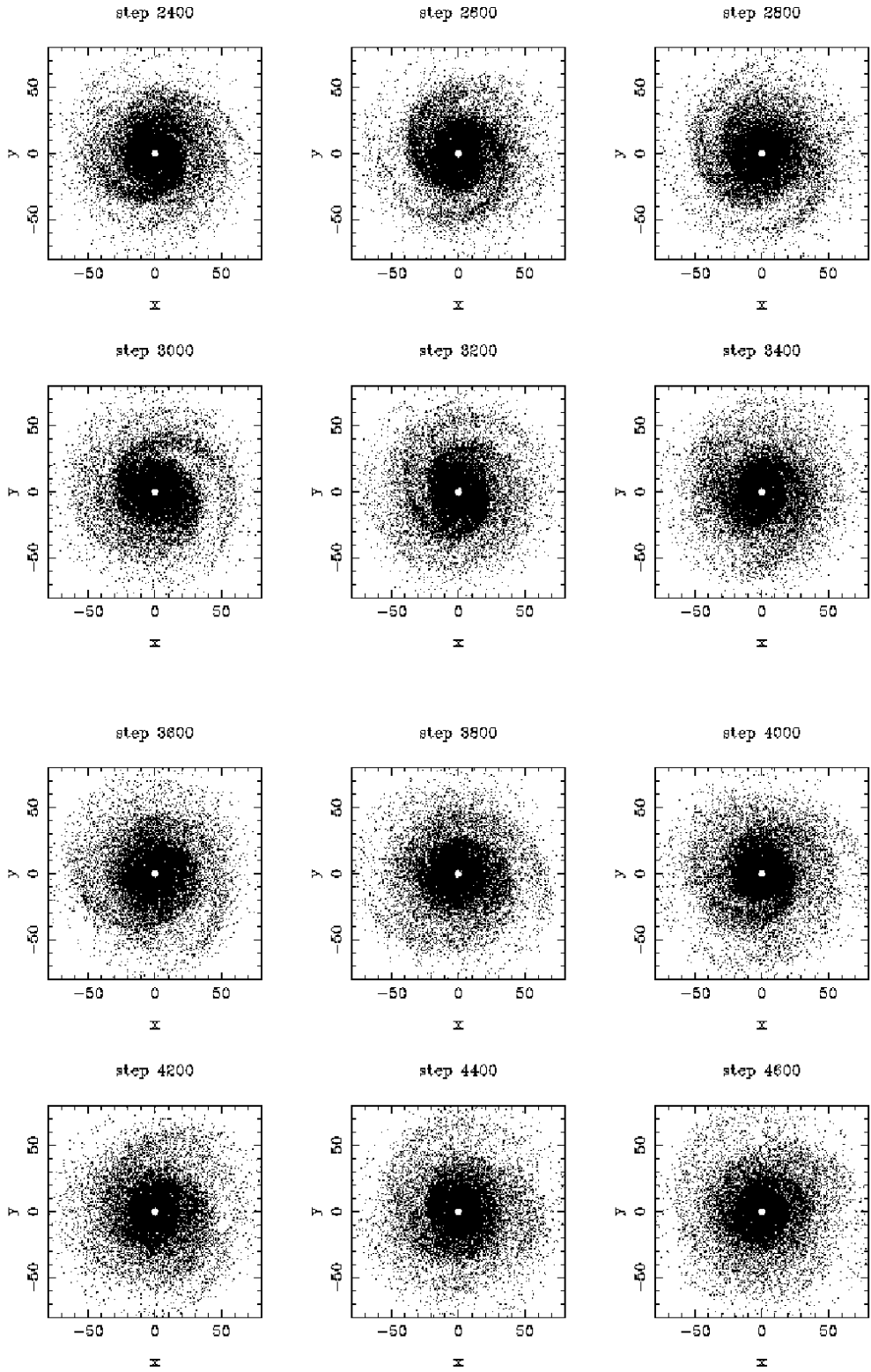

\begin{center}
\epsfysize=3.8in
\epsffile{figure1c.epsi}
\end{center}
\smallskip
\begin{center}
\epsfysize=3.8in
\epsffile{figure1d.epsi}
\bigskip
\caption{ (continued).}
\end{center}
\end{figure*}

Such long-lived spiral patterns are related to the
spontaneously growing {\em modes} in the galactic resonant cavity
rather than {\em wave trains}.
Furthermore, the spiral modes cannot grow indefinitely.
The wave amplitude has to be clamped at a finite value as is
observed in real galaxies.  The two conditions, i.e., the
continuous outward angular momentum transport by a long-lived
spiral pattern coupled with a finite wave amplitude, imply
that at the quasi-steady state of the wave mode
the outward-transported angular momentum cannot come from
the wave itself (for otherwise the wave amplitude will continue
growing without bound), but has to come from the basic state of
the galactic disk, since the wave and the basic state are the
only two subsystems that we divide the disk into.
The secular evolution of the mass distribution of the galactic
disk is thus an inevitable consequence of the requirement of
global angular momentum conservation and the assumption of a
quasi-stationary spiral structure; or, to put it in another way,
the globally self-consistent quasi-steady spiral solution
is maintained at the expense of a continuous secular basic state evolution.
The source and sink of the angular momentum transported 
by a quasi-stationary spiral mode both reside in the
basic state: they are the inner and outer disk, respectively
(the dividing line between the loading and unloading of the
angular momemtum is at the corotation radius, as we will
discuss below).  The energy and angular momentum exchange between 
the wave and the basic state thus serves as a damping mechanism 
for the spontaneously growing density wave mode:
since the wave has negative energy and angular momentum density
inside corotation, to receive energy and angular momentum from
the basic state in the inner disk limits the wave growth; similarly,
since the wave has positive energy and angular momentum denity outside
corotation, to dump energy and angular momentum to the basic state
also limit the wave growth in the outer disk.  In the end
the nonlinearity in this exchange process helps to clamp the
wave amplitude at a particular value which is mainly determined by the 
basic state properties (Zhang 1998).

What then is the mechanism through which the outward-transported
energy and angular momentum are loaded onto the density wave in 
the inner disk, as well as unloaded in the outer disk?
The relevant mechanism obviously has to involve the interaction
of the basic state and the wave mode: Specifically, it has to
involve a dissipative energy and angular momentum exchange
between the wave and the basic state; i.e., the loading of angular momentum
onto the wave from the basic state inside corotation, and unloading
of this angular momentum outside corotation, with the wave itself
being the carrier for the angular momentum transport.
Such a mechanism for wave/basic state interaction was indeed found
(Zhang 1996,1998,1999), and we summarize the essential
characteristics of this mechanism below.

In was first shown in Zhang (1996) that for a self-sustained
spiral mode, the minimum of the gravitational potential of
a spiral density wave lags behind the maximum in density in the
azimuthal direction inside corotation, and vice versa outside corotation.
The phase shift between the potential
and density spirals means that there is a torque exerted by the
potential spiral on the density spiral, and, at the quasi-steady
state of the wave mode, a secular transfer of energy and
angular momentum between the disk matter and the density
wave.  The existence of the phase shift between the potential and
density spirals of a self-sustained spiral mode is partly a result of
the long range nature of gravitational interaction.  It is for this
reason that a skewed bar or other skewed large scale patterns
will also possess a phase shift and the associated collective
dissipation through essentially the same mechanism.

The torque T(r) applied by the spiral potential on the disk density
in an annulus of unit width can be written as (Zhang 1996, 1998)
$$
T (r) = dL/dt
= r
\int_0^{2  \pi}
- \Sigma (r \times \nabla {\cal{V}})_z d \phi
$$
\begin{equation}
=- \pi m r \Sigma_1(r) {\cal{V}}_1(r) \cdot \sin (m \phi_0(r))
,
\label{eq:torque}
\end{equation}
where $\Sigma$, ${\cal{V}}$, $\Sigma_1$, ${\cal{V}}_1$
are the disk surface density, potential,
the spiral perturbation density and spiral perturbation
potential in the annulus, respectively;
$L$ is the angular momentum of the disk matter in the annulus,
$\phi_0$ is the potential-density phase shift, which is
found to be positive
inside corotation (potential lags density) and negative
outside corotation (potential leads density), and $m$ is the number of
spiral arms.  It can be seen from (\ref{eq:torque}) that the torque $T(r)$ is
non-zero only when the phase shift $\phi_0$ is non-zero.
Furthermore, the contributions of both the
gravitational and advective torque couplings are included in the
single torque integral given in equation (\ref{eq:torque}) at the
quasi-steady state of the wave mode.  The proof of this fact
is given in the Appendices of Zhang (1998,1999).

The energy and angular momentum exchange between
the disk matter and the density wave at the quasi-steady state
as indicated by equation (\ref{eq:torque})
is achieved through a temporary local gravitational instability
at the spiral arms (Zhang 1996).  The length scale of this instability
at the solar radius is calculated to be about 1 kpc,
comparable to the length scale of the giant HI and
molecular complexes near the Galactic spiral arm region (Elmegreen 1979).

In Figure 3, we show an image of an N-body
spiral mode from Zhang (1996).  There we see the density enhancement
at the inner edge of the spiral pattern inside corotation (the
corotation radius $r_{co}= 30$ in this case), reminiscent of the 
dust lanes observed at the leading edges of the spiral arms of physical 
galaxies which signal the presense of gaseous shocks.  However, here in a 
collisionless particle disk of the N-body simulation
we have obtained the signature of a shock wave, which is a
phenomenon generally attributed to a dissipative system.  In what 
follows we present further evidence that a spiral density wave is 
in fact a propagating front of collisionless shock (Zhang 1996).

\bigskip
\begin{figure}[!htbp]
\begin{center}
\epsfysize=3.0in
\epsffile{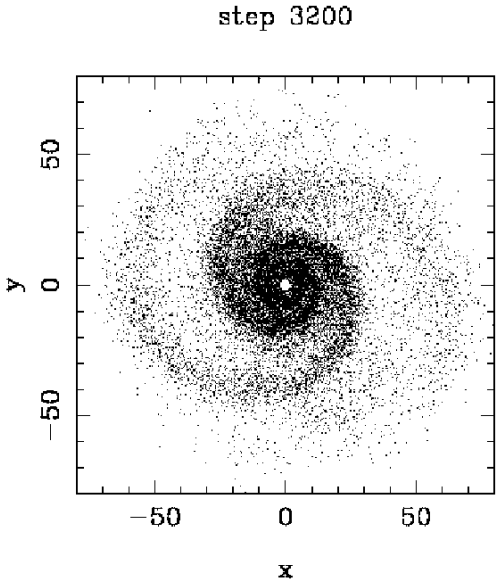}
\bigskip
\caption{Detailed morphology of a spontaneously-formed
N-body spiral mode, showing the density maxima at the leading edge of the
spiral pattern for locations inside corotation (Zhang 1996).}
\end{center}
\end{figure}

In Figure 4, the azimuthal variations of the different disk parameters
are plotted.  We see from (a) that the potential indeed lags the density 
for this typical radial location inside corotation.
Furthermore, from (c) it is seen that Toomre's Q parameter has
a clear minimum in the higher density region of the spiral arms.
This sudden drop in Q signals the presence of local gravitational instability
at the arm region.  (d) shows that the velocity component perpendicular 
to the spiral arm suffers a sharp jump from supersonic to subsonic 
(the average sound velocity is about 0.04, as shown in (b)), further
reinforcing the impression of the presence of a shock.

\begin{figure*}[!htbp]
\begin{center}
\epsfysize=3.8in
\epsffile{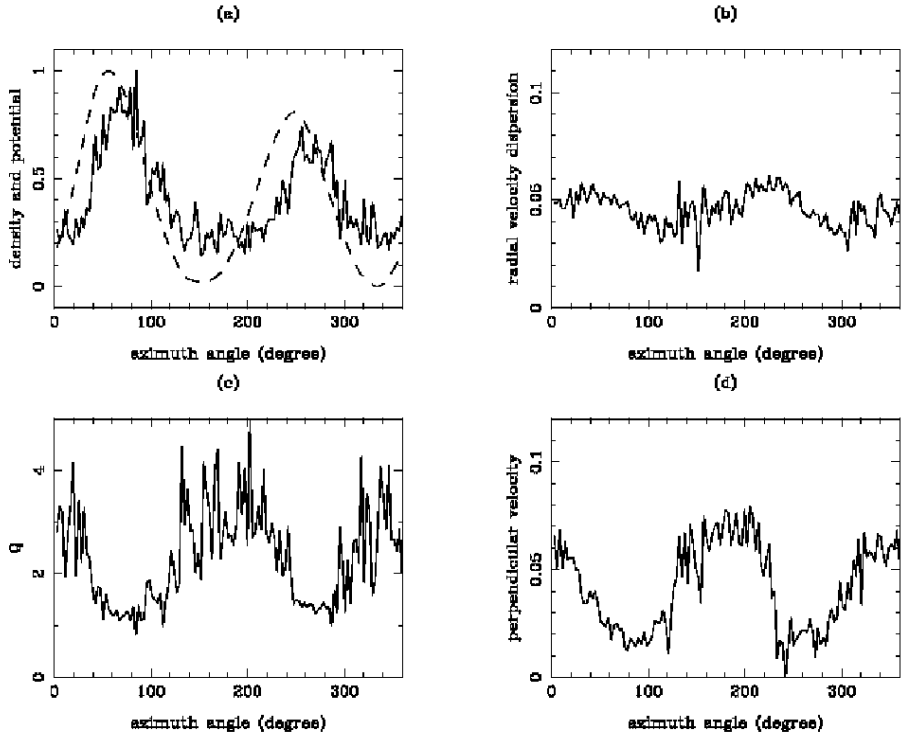}
\bigskip
\caption{Spiral gravitational shock. Different
frames show the azimuthal distributions of the following parameters:
(a) Surface density (solid line) and negative potential (dashed line).
(b) Radial velocity dispersion.
(c) Toomre's Q parameter.
(d) Velocity component perpendicular to the spiral arm.
The above quantities are computed at a radius of 14.5
(From Zhang 1996).
}
\end{center}
\end{figure*}

The gravitational instability and the associated small-angle
scattering of the streaming stars at the arms of a self-sustained
spiral wave is what breaks the conservation of the Jacobi for 
a single stellar orbit in a smooth and steady-amplitude spiral potential,
or equivalently the no-wave-basic-state interactionc conclusion of
LBK, and this allows the stellar orbit to display secular decay or increase.  

\begin{figure*}[!htbp]
\begin{center}
\epsfysize=2.7in
\epsffile{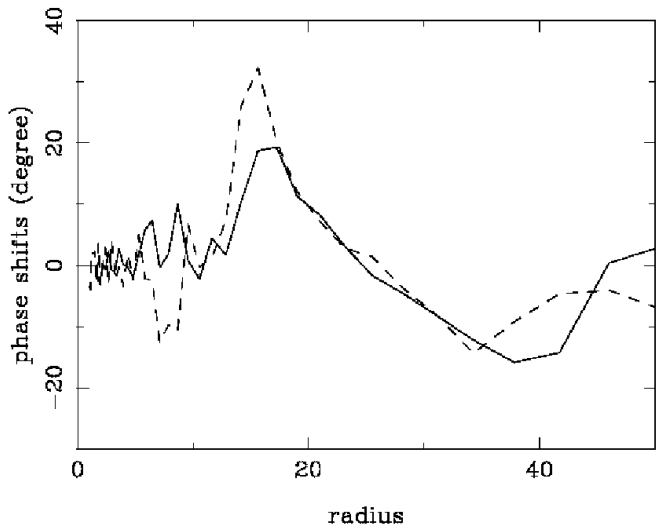}
\bigskip
\caption{Potential and density phase shift for the
stellar component (solid) and the gas component (dash),
respectively, for the spiral mode in the two-component
N-body simulations of Zhang (1998) at time step 1600.}
\end{center}
\end{figure*}

Essentially the same result can be obtained for a star-gas
two-component disk (Zhang 1998).
In the past, discussions of secular evolution in galaxies have
focused on the accretion of gas under the influence of a central
bar.  This originates partly from the mis-conception
that ``gas is dissipative, whereas stars are not''.
However, as is well known, the microscopic
viscosity in the gas component is inadequate to support
a reasonable accretion rate even for proto-stellar accretion
disks (see, e.g. Pringle 1981).  Instead, the
gravitational viscosity due to the collective dissipation
effect of the non-axisymmetric large-scale structures has to be
responsible even for the accretion of the gas component. 
This is because gravity does
really distinguish whether the underlying matter is made of stars or gas.
The past star-gas two-component N-body simulations 
have often found that the phase shift between the stellar and gaseous 
densities are usually small (Carlberg \& Freedman 1985),
especially incomparison with the phaseshifts of these densites
with respect to their common spiral potential (Figure 5).   
These phaseshifts causes stars and gas to both drift 
towards the center as well as being heated. 
We will calculate these spiral induced evolution 
rates quantitatively in the next subsection.

\subsection{Astrophysical Consequences}

As a result of the wave-basic state angular momentum exchange,
an average orbiting star in the basic state inside corotation
loses energy and angular momentum to the wave secularly
and tends to spiral inward (Figure 6).  Similarly, a star outside 
corotation gains energy and angular momentum from the wave and
drifts outward secularly (Figure 7).
The mean orbital radius evolution leads to a corresponding
disk surface density evolution which we plot in Figure 8, where
the dashed line in each frame indicates the surface density at the
earlier time step, and the solid line the later time step. 
Since the star inside corotation drifts inward, and outside corotation
drifts outward, at corotation the surface density decreases with
time.  Note that as a galaxy evolves, the spiral pattern
speed tends to decrease and thus the corotation radius tends to
move outward (Toomre 1981). 
We also see the trend of increasing disk central density
together with the build-up of the extended outer envelope,
as predicted.  

\begin{figure}[!htbp]
\begin{center}
\epsfysize=2.0in
\epsffile{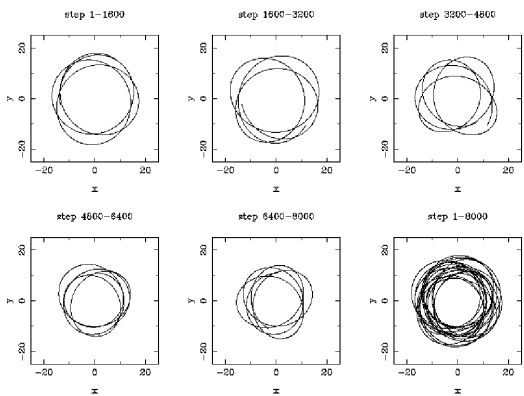}
\bigskip
\caption{Evolution of orbit trajectory
for a typical star inside corotation. From Zhang (1996).}
\end{center}
\end{figure}

\begin{figure}[!htbp]
\begin{center}
\epsfysize=2.0in
\epsffile{outsiderco.epsi}
\bigskip
\caption{Evolution of orbit trajectory
for a typical star outside corotation. From Zhang (1996)}
\end{center}
\end{figure}

\begin{figure}[!htbp]
\begin{center}
\epsfysize=2.0in
\epsffile{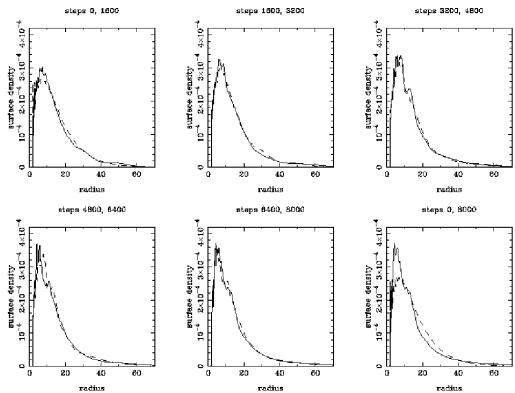}
\bigskip
\caption{Evolutuon of disk surface density.  From Zhang (1996).}
\end{center}
\end{figure}

A quantitative estimate of the rate of orbital change can be obtained from
equation (\ref{eq:torque}), which turns out to be
(Zhang 1998)
\begin{equation}
{{dr} \over {dt}}
= - { 1 \over 2} F^2 v_0 \tan (i) \sin(m \phi_0)
,
\label{eq:decay}
\end{equation}
where $F$ is the fractional wave amplitude, $v_0$ is the circular
velocity of the star, $i$ is the pitch angle of the spiral, and
$m$ and $\phi_0$ are again the number of spiral arms and the
potential-density phase shift, respectively.

To calculate the secular evolution rate for our own Galaxy,
we assume a two-armed spiral pattern of 20\% amplitude 
and $20^{o}$ pitch angle (Drimmel 1991), which is appropriate for 
the average Hubble type that our Galaxy had during the past $10^{10}$ 
years of evolution.  This set of values gives an orbital
decay rate of 2 kpc per Hubble time using (\ref{eq:decay});
the same set of spiral parameters also nicely fitted
the observed age-velocity dispersion curve (Figure 1) using
the velocity diffusion equation we will derive later,
increasing the credibility of this particular set of choice
of spiral parameters.

Therefore, a star in the Sun's orbit will not make it all the way 
in to the inner Galaxy in a Hubble time.
However, the corresponding mass accretion rate across any 
Galactic radius inside corotation is given by
\begin{equation}
{{dM} \over {dt}} = 2 \pi r {{dr} \over {dt}} \Sigma
,
\label{eq:dm}
\end{equation}
where $\Sigma$ is the disk surface density.
Using (\ref{eq:dm}),
and a solar neighborhood average disk surface density of
$60 M_{\odot} pc^{-2}$  (Bahcall 1984; Kuijken \& Gilmore 1989),
the mass accretion rate for the
Galaxy disk is found to be about $6 \times 10^9 M_{\odot}$ per $10^{10}$
yr.  A substantial fraction of the Galactic bulge can thus be built up 
in a Hubble time.  The vertical drift (or velocity dispersion
increase) needed for a star to truly become a bulge
star is produced by the isotropic heating effect accompanying
accretion, which we will discuss a little later.  

Observationally, bulges are found to be old, and encompass a wide range
of metallicities (see, e.g. Goudfrooij, Gorgas \& Jablonka 1999),
with a clear radial gradient in both the age and metalicity distribution
which appears to be the continuation of a similar gradient 
in the disk (Courteau et al. 1996).
The secular evolution process discussed in this work
predicts that the abundance gradient will be enhanced  
with time, since it is effectively an inner disk contraction process 
which amounts to about 2 kpc of disk scalelength reduction in a Hubble time.  
This, coupled with the shorter dynamical time scale in the inner 
disk and the resulting enhanced element recycling
rate, leads to an increase of the abundance gradient with time.
This is in contrast with the gaseous-bar secular evolution scenario
(e.g. Friedli, Benz \& Kennicutt 1994 and the references therein) 
which generally predicts a flattening of the abundance gradient due 
to gas inflow.  Gas accretion also tends to create a bluing of the
nuclear region even though early type bulges are often observed to be red.
Furthermore, while gas accretion could be 
important for the morphological transformation among later Hubble
types, it is insufficient to explain the transformation of Sb to Sa,
and Sa to S0, etc..  Bulge formation also cannot be solely due to
the dissolution of a pre-exsiting stellar bar as some theories
suggest since observationally the bulge light is found
to be added onto the disk light, instead of substrated from it
(Wyse, Gilmore,\& Franx 1997).
Part of the reasons previous secular evolution theories
arrived at the above results was due to the unequal treatment of stars and gas,
i.e. the relevant numerical simulations usually considered the {\em response} 
of the gas under the {\em applied} stellar bar potential; the viscosity 
of the gas in these calculations were also introduced artificially 
and its value somewhat arbitrarily.

Another important consequence of spiral-induced wave-basic state
interaction is the secular heating of the disk stars,
believed to be the main process responsible for producing the
age-velocity dispersion relation of the solar neighborhood stars 
(Figure 1).  As we have mentioned above, the secular heating process
allows the stars to gradually drift out of the galactic plane 
as they spiral inward, and eventually become bulge stars.

The secular heating of the disk stars works as follows.
Since a spiral density wave can only gain energy and angular momentum
in proportion to $\Omega_p$, the pattern speed of the wave,
and a disk star which moves on a nearly circular
orbit loses its orbital energy and angular momentum
in proportion to $\Omega$, the circular speed of the star,
an average star cannot lose the orbital energy entirely to the wave;
thus, the excess energy serves to heat the star when it crosses
the spiral arm.  For our Galaxy,
the diffusion coefficient due to the spiral-induced secular heating is
estimated to be
\begin{equation}
D = (\Omega - \Omega_p) F^2 v_c^2 \tan (i) \sin(m \phi_0 ) \approx
6.0 (km s^{-1})^2 yr^{-1},
\label{eq:D}
\end{equation}
if using the same set of spiral parameters
as used above for estimating Bulge building (i.e. a 20$^o$ pitch angle
and 20\% amplitude two armed spiral).
This value of D fits very well the age-velocity
dispersion relation for the solar neighborhood stars as can
be seen in Figure 1.
The above expression for D can be shown to be approximately
independent of galactic radius (Zhang 1999), which would
reproduce the observed isothermal distribution of the stellar and gaseous 
mass across the Galaxy (Gilmore, King \& van der Kruit 1990).
Since the spiral gravitational instability which mediates the
wave-star energy and angular momentum exchange is a local
instability, the heating of the disk stars is approximately 
isotropic, and all three dimensions of the velocity dispersion
increases at approximately the same rate as is observed (Wielen 1977).

The velocity dispersion of the gas in the high redshift Damped $L_{\alpha}$ 
systems (DLAs), which are believed to be the candidate primordial disk galaxies,
is found to be around 10 km/s (Wolfe 2001).  This can be gradually increased 
to the stellar velocity dispersion of 40 km/s of the thick disk stars of 
the present-day Milky Way-like galaxy through the above spiral 
heating mechanism.  The metalicity of the DLA systems
are found to have evolved little between z=2-4 (Prochaska, Wolfe \& Gawiser
2000), consistent with the fact that the disks during this period have not 
formed prominent spiral patterns and thus the metal enrichment evolution 
is not prominent.  Emergence of the spiral heats the disk immediately,
as confirmed in the N-body simulations.  This can be a natural explanation
for the discontinuity found in the age-velocity dispersion relation
of the solar neighborhood stars 11 Gyr ago.

A similar energy injection into the interstellar medium can serve
as the top-level energy source to power the subsequent supersonic 
turbulence cascade (Zhang 2002; Zhang et al. 2001), which naturally
explains the size-linewidth relation of the interstellar clouds.  

We thus see that the spiral and bar-induced radial mass accretion process
leads to the building up of the bulge, and causes the Hubble type of a galaxy 
to evolve from late to early.  Such morphological transformation is 
observationally most pronounced in dense BO clusters, though its less 
pronounced counterpart in the field has also been observed
(Lilly et al. 1998).  The enhanced mass accretion rate for
cluster galaxies is in part due to the large amplitude and open spiral patterns
induced through tidal interactions of neighboring galaxies,
since the effective evolution rate is proportional to wave amplitude 
squared and the spiral pitch angle squared (equation 2) 
(note that the phase shift $\phi_0$ itself is approximately
proportional to spiral pitch angle).
Preliminary evidence has been found that the brightness (or amplitude)
of the spiral pattern is signiicantly hight for cluster galaxies 
compared to the field galaxies of the same rotation curve and size
(Aragon-Salamanca et al 2002). 
Tidal interactions among neighboring galaxies have also been found to
produce enhanced disk mass accretion and nuclear activity (Byrd et al. 1986;
Zhang, Wright, \& Alexander 1993).
The necessity of invoking environmental effects to enhance the strength
of the spiral structure does not make the process we discuss less
interesting or relevant.  Just as in the case of the development of
a human being, even though the environmental input and nourishment are
important, without an innate mechanism for development and maturation,
the human growth process would not happen. 

\subsection{Formation of Coherent Patterns in Disk Galaxies as
an Example of Non-Equilibrium Phase Transitions}

It is well known that for an isolated system, the direction of
entropy evolution is towards an increasing degree of macroscopic 
uniformity, corresponding to increasing entropy.
For open systems at far-from-equilibrium conditions, however,
it often happens that the usual near-equilibrium
thermodynamic branch of the
solution becomes unstable, and new types of highly organized
spatial-temporal structures emerge spontaneously.
Due to its similarity to equilibrium phase transitions,
this kind of spontaneous structure formation in nonequilibrium
systems has been termed ``nonequilibrium phase transitions'',
and the structures thus formed ``dissipative structures''
(Glansdorff \& Prigogine 1977; Nicolis \& Prigogine 1977)
to emphasize the constructive role of dissipation in the
maintenance of these nonequilibrium structures.

The large-scale coherent patterns formed in open and nonequilibrium 
systems are functional as well as architectural.
One of the important functions of these ``dissipative structures''
is to greatly accelerate the speed of entropy evolution of these
systems towards reaching thermodynamic equilibrium, or at least
reducing the degree of nonequilibrium.
The local highly-ordered structure (which has low entropy)
maintains its constant entropy in the meta-stable state
by continuously exporting the entropy it produces to its environment. As
a result, the entropy of the structure plus the environment increases
at a much faster rate than when the system was
still on the thermodynamic branch of the solution.

The spiral (or bar) patterns of galaxies have many characteristics 
of a typical ``dissipative structure''.  First, as
we have shown, a quasi-stationary spiral mode is maintained
by the opposing effect of the
spontaneous growth tendency and local dissipation,
with a continuous flux of energy, angular momentum and entropy
through the system carried by the trailing spiral wave itself.
Second, it can be shown that the formation
of spiral structure accelerates the speed of entropy evolution
of a spiral galaxy, compared to that of a uniform disk, by
several orders of magnitude (Zhang 1992).
Thirdly, since a spiral mode is a global
instability in the underlying basic state of the disk,
the spontaneous emergence of the spiral pattern (which is obviously
a global symmetry-breaking process)
happens as long as the disk satisfies certain far-from-equilibrium
constraints (i.e. the basic state characteristics must allow the linear
growth rate of a spiral mode to be greater than zero).
Lastly, the characteristics of the quasi-stationary spiral pattern
formed are determined solely
by the properties of the basic state,
and not by the accidentals of the
external perturbations.
This last point is reinforced
by the N-body simulations of
tidal spiral patterns in slightly unstable disks, where it was found that
after the initial transient state,
the characteristics of the tidally-induced patterns
correlate strongly with
the properties of the basic state, rather than with the nature of
the encounter (Donner \& Thomasson 1994).
These characteristics of the spiral structure
clearly identify it as an example of a
``dissipative structure'' defined by
Glansdorff \& Prigogine (1971), and the spontaneous
formation and stabilization
of a large-scale spiral
mode as an example of a nonequilibrium phase transition.

\section{COMPARISON OF THE HIERARCHICAL CLUSTERING
AND THE SECULAR EVOLUTION PREDICTIONS}

Currently the working paradigm for the formation and evolution
of galaxies and structures is the hierarchical clustering or 
cold dark matter (CDM) model.  It has demonstrated successes in
reproducing the angular spectrum of the cosmic microwave background 
(CMB) radiation as well as many of the aspects of
the distribution of large scale structure
(see, e.g., Bahcall et al. 1999 and the references therein).
However, at the individual galaxy level, the CDM model has encountered
serious challenges in attempting to reproduce the observed galaxy
properties.  The standard CDM is now replaced by the $\Lambda$CDM paradigm, 
though many of the problems still remain.  

In a recent article, Peeble (2002a) compared the current state of
cosmology with the state of physics at the turn of the 19th/20th
century, and commented that several known problems of
the CDM could potentially turn out to be the same type of ``Kelvin-level
clouds'' which a century ago resulted in the revolution of modern physics,
i.e. the creation of relativity and quantum mechanics theories.  
These problems include ``the prediction that elliptical galaxies form by 
mergers at modest redshifts, which seems to be at odds with the observation
of massive quasars at z $\sim$ 6; the prediction of appreciable
debris in the voids defined by $L_*$ galaxies, which seems to be at odds
with the observation that dwarf, irregular, and $L_*$ galaxies share
quite similar distributions; and the prediction of cusp-like
dark matter cores in low surface brightness galaxies, which is at
odds with what is observed'' (Peebles 2002a).

Historically, CDM type of theories were invented partly to get around 
the problem that there does not seem to be sufficient time for the
seeds of the anisotropies observed on the cosmic microwave background,
about one part in $10^5$, to grow into the nonlinear 
structures we see today by gravitational means alone, which requires
seeds of one part in $\sim 10^3$ at the time of recombination (z=1000). 
Furthermore, the Big Bang nucleosysthesis model also requires a
significant amount of non-baryonic dark matter (Primack 1999) if the
universe is flat as the inflation scenario suggests.  

Given the partially {\em ad hoc} nature of the introduction of the CDM 
(especially since after 30 years of search, no evidence of the 
existence of the CDM material has been found), it should not come 
as a total surprise that problems surface when observational data 
become available to allow a detailed comparison
with the predictions of CDM model.  In fact, most of the problems
of the CDM scenario (which Peebles had quoted three above, and which
we will list several more in the following) can be characterized
by that it prescribes a medium for structure formation which
is too clumpy (or can easily become too clumpy)
on small scales, yet too smooth on large scales.
For example, the cusp problem and the satellite abundance problem
are all both due to the over-clumpiness of the medium on small scales,
so is the problem of small disks or rapid angular momentum loss
during disk formation (White \& Frenk 1991); on the other
hand, the over-smoothness of the medium on large scales underlies
the problem of its inability to account for early quasar formation,
the early formation of giant high redshift clusters (Francis
et al. 1997; Steidel et al. 1998; Williger et al. 2002),
as well as the problem of accounting for the observed bubble and 
void appearance of large-scale structure (Geller and Huchra 1984).
Furthermore, in a purely bottom-up structure formation scenario
such as the CDM, it is very difficult to account for the alignment
of the spin axis of the bright galaxies in a cluster (Ozernoy 1994a
and the references therein; Kim 2001), as well as the observed
galaxy-cluster-supercluster alignment effect on large scale
(West 2001).

In what follows we contrast a number of the major 
predictions of the hierarchical clustering (CDM)/merger
scenario with that of the secular evolution (SE), focusing on individual
galaxy properties, and compare both predictions with the known 
observational facts when available.  Through this cross comparison,
we wish to demonstrate that secular evolution is indeed a much
more natural paradigm in explaining the properties
(as well the evolution of these properties) of the
observed galaxies:
\begin{itemize}
\item The CDM model
predicts that the total number of galaxies of all Hubble types
per comoving volume should decrease with time due to merger events;
whereas SE predicts that the comoving number density of
all Hubble types should remain nearly constant,  
and the number counts for individual Hubble types should evolve 
according to the morphological transformation picture.
Recent large surveys such as the Caltech Faint 
Galaxy Redshift Survey (CFGRS) and Slone Digical Sky Survey (SDSS)
have shown that there is essentially no evolution of the
total number density of galaxies per co-moving volume between z=0 and 1
(Cohen 2002; Yasuda et al. 2001).
\item According to the CDM paradigm the field galaxies should evolve faster
through merger process (since merger is known to happen more frequently
in the fields), and rarely happens in clusters due to the 
relatively high speed nature of cluster galaxy encounters.  
SE on the other hand predicts that cluster galaxies should 
show a faster secular evolution 
rate than field galaxies due to the tidal-interaction-enhanced 
and spiral-mediated mass redistribution.  The well-known Butcher Oemler
effect for cluster galaxies (Butcher \& Oemler 1978a,b)
as well as the observation of field galaxies (Ellis 1997)
indicate that the cluster galaxies have a much large
morphological evolution rate than field galaxies.
\item In the CDM paradigm, there is no direct relation between the
kinematics and energetics of the stars and gas in a given spiral
disk;  whereas the SE theory predicts that stars and
gas should appear on the same energetic hierarchy
at the 1 kpc spatial scale (Zhang 2002), 
because the interstellar medium receives similar amount 
of energy injection per unit mass from the spiral density wave as the 
stellar component.  The observations indicate a clear correlation
of stellar and gaseous kinematics on 1 kpc spatial
scale (Larson 1979, 1981; Fleck 1982).
\item The CDM galaxy formation model (Kauffmann 1996) 
prescribes that the collapse of gaseous material within the
clustering dark matter halo produces disks,
mergers of nearly equal mass disks produce ellipticals, and
ellipticals subsequently grow disks if left undisturbed.
One consequence of this prescription is that late-type spirals,
which have a large disk-to-bulge ratio, should have older
bulges than do early-type spirals, since to have a larger disk
the galaxy must have been undisturbed and be able to accrete gas
for a longer time, contrary to the observed trend
(Wyse et al. 1997); the SE theory on the other hand predicts
that the early type bulges are older since they have 
accreted mass for a longer time and most of the accreted mass
is in stellar form from the local vicinity, since the central region formed
stars early due to the deeper potential well and shorter dynamical
time.
\item The CDM theory does not predict any correlation 
of the disk heating time scale and the angular momentum transport 
time scale; whereas in the SE scenario these 
two time scales are found to be tightly correlated 
(see equations \ref{eq:decay} and \ref{eq:D}).  The observed
continuous change of bulge to disk ratio and other type of
bulge-disk connections (Courteau et al. 1996)
attest to the correlation of heating and angular momentum transport
time scales during the morphological evolution of the disks.
\item The disk galaxies produced in the CDM simulations are much 
too small and rotate much too fast compared to the observed 
galaxy disks (White \& Frenk 1991; Navarro \& Steinmetz 2000).  
This problem is made especially acute by the observational fact 
that the observed damped $L_{\alpha}$ systems at high z (Wolfe 2001), 
which are the most likely candidates for primordial disks, 
are usually quite large.  SE starts from the 
initial condition of large disks at high z, which become
the progenitors of today's large early type galaxies.
\item The halo merging scenario tends to create a cuspy mass
distribution in the center of a proto galaxy.  The observed
young galaxy candidates in the nearby universe, i.e,. the low surface
brightness (LSB) galaxies (Impey \& Bothun 1997) and dwarf galaxies 
are found not to possess such cuspy cores (de Blok, McGaugh, \& Rubin
2001).  Even invoking maximum feedback 
does not solve the cusp problem (Gnedin \& Zhao 2002).
The SE scenario on the other hand
starts from the more flattened morphological distribution of
a LSB disk, and the central density of a galaxy is increased 
gradually through a Hubble time of evolution.
\item The CDM simulations of galaxy formation have not 
been able to fit simultaneously the observed zero
point of the Tully-Fisher (TF) relation and the local
luminosity function, a problem related to the
small-disk problem mentioned above (White \& Frenk 1991).
Furthermore, the CDM model naturally predicts that the luminosity 
of an individual galaxy is proportional to the 
third power of its circular velocity (White 1997;
Dalcanton, Spergel \& Summers 1997);
To arrive at the observed fourth power TF relation requires fine-tuning
of the feedback and cooling parameters (van den Bosch 2000).
The SE scenario on the other hand predicts that the
entire Hubble sequence from the spiral disks to disky ellipticals 
should follow roughly the same Tully-Fisher/Faber Jackson relation 
(Zhang 1999).  No modified Newtonian dynamics is needed to explain the
fact that LSB disks fall onto the same TF relation as 
normal spirals, as long as the decrease in surface brightness
of a LSB galaxy is compensated by the increase of mass-to-light
ratio in the usual virial theorem type of derivation of TF relation.
\item In the CDM paradigm, large galaxies are formed
out of the mergers of smaller ones, and therefore should form last.
Recent studies of high z galaxies have shown that 
exactly the opposite is observed: i.e., the larger the mass of 
a galaxy, the shorter the time scale of its formation (Thomas,
Maraston \& Bender 2002; Boissier et al. 2001).
In the SE scenario large galaxies tend to form
quicker because of the more rapid gravitational collapse to
form disks and the faster rate of evolution due to a more
prominent spiral structure on the massive disk.
\item The observed elliptical galaxies come in two types. While
the boxy giant ellipticals have characteristics which indicate that
they are likely to be the product of mergers, i.e., being pressure
supported, having multiple nuclei, having two populations of globular 
clusters of different colors, and are radio and X-ray
loud, etc., the more numerous disky type ellipticals on the other
hand are mostly rotationally supported, show little evidence of mergers,
having only one globular cluster population, and being radio and
X-day quiet (Zhang 1999 and the references therein; Lee 2002).  
It is our belief that this dichotomy of characteristics is due to that
boxy ellipticals are the true merger products, whereas
disky ellipticals are produced mostly by secular evoltion,
\item N-body simulations have shown that stellar mergers tend to flatten
out the abundance gradient from that of the observed power law shape
(Mihos \& Hernquist 1994a).  Gas rich mergers (i.e., 10\% gas) tend to 
create a distinctive dense core (Mihos \& Hernquist 1994b) which is once again 
not observed in normal ellipticals.  The multi-component merger scenario
(Weil \& Hernquist 1994, 1996) is not relevant to the explanation 
of the gradual decrease of disk galaxy population with 
decreasing redshift.  The observed density and abundance profiles
can be naturally produced and enhanced through secular evolution process.
\end{itemize}

The above comparison, as well as the fact (see, e.g. Peebles 2002b
and the references therein) that it appears difficult
to reproduce simultaneously the spectrum of $L_{\alpha}$ forest 
(which requires plenty of small-scale CDM power) and the individual galaxy 
properties (which are troubled by too much small scale power)
indicate that the true underlying structure formation theory
may not be purely gravitational, but in addition may include other
nonlinear processes such as primordial turbulence (von Weisacker 1951;
Gamov 1952; Ozernoy 1974b).  The supersonic shocks associated with turbulence
can solve the time-of-growth problem of the high-z quasars and high
z ellipticals without the need 
of dark matter (though it does not necessarily exclude it); it also
naturally produces the bubble and void appearance of the large
scale distribution.  Turbulence cascade can produces the observed
scale-invariant mass spectrum (Ozernoy 1974b) just as the CDM theories.
It also allow small scale structure
to be present but not condense and collapse to form cusps and cores.
The problem with the primordial turbulence scenario 
is of course why such supersonic motion has not left significant
imprints on the cosmic microwave background: the observed CMB
fluctuation of $10^{-5}$ does not leave room for significant
velocity fluctuation at the recombination time
if the turbulent matter was coupled to radiation then.
Recently, the first evidence of primordial turbulence in the 
form of a Kolmogorov scaling relation between temperature increment 
and angular separation on the CMB has been detected
(Bershadskii \& Screenivasan 2002).  The issue of whether this
indicates a real turbulent state of the matter at decoupling,
or else it is rather the fossil of an earlier turbulent stage
before decoupling (Gibson 2000) is not yet clear.

\section{FUTURE RESEARCH}

Even though the analytically derived evolution rates (\ref{eq:decay})
(\ref{eq:D}) indicate that bulge formation through secular
evolution is realistic if one adopts a realistic
set of spiral parameters for pitch angle and wave amplitude,
the past 2D N-body simulations have often produced much smaller
spiral amplitude compared to those actually observed in physical galaxies
and thus a relatively low evolution rate, especially
when the adopted (bulge+halo) mass to disk mass ratio is high 
(a comparison of the 2D N-body simulation results of Zhang
1996 and Zhang 1999 shows clearly the effect of the
spheroidal-to-disk ratio to the evolution rates obtained).
This result is believed to be partly an artifact due
to the enforcement of a rigid spheroidal component in the 2D simulations.
Recent studies by Athanassoula (2002) found that making the halo active
leads to enhanced bar formation. It is thus expected that an enhanced spiral
formation and enhanced secular evolution rate will also be obtained
in a full 3D and live halo simulation of spiral disks.  
Placing a galaxy under the tidal influence of neighboring
galaxies in realistic group or cluster environment will also help
to obtain a larger spiral amplitude and thus a higher evolution rate.
Another issue which can be explored by a 3D simulation is the 
secular evolution caused by a skewed mass distribution such as found 
in the high-z field galaxies in the Hubble Deep Fields. 
These skewed 3D mass distribution should produce the same kind
of phase-shift and torque relations as in spiral galaxies.
The secular evolution produced by
these structures may also played a role for the direct formation
of some high-z elliptical galaxies (Peebles 2002c and the references
therein) without going through the disk formation phase.

The secular evolution theory we presented here
describes a morphological transformation process of individual galaxies.
By itself it does not uniquely specify a cosmology,
though it does hint at the elements and consequences 
of such a cosmology.  For example, it favors large disks to form early, 
and subsequently undergo morphological transformation mainly due to 
the mediation of global structures such as spirals, bars, and 
three dimensional twisted isophotes.  Environmental effect
accelerates the evolution speed, but it operates mainly through the
mediation of internal global structure, and not through the actual
merging of galaxies.  As we gradually elliminate
uncertain elements from our knowledge of the galaxy formation
process we will be able to better constrain the 
elements of earlier cosmological processes.

\vskip 0.4cm

I would like to express my sincere gratitude to professor Hong Bae Ann, 
professor Hyung Mok Lee, as well as the Local Organizing Committee 
for inviting me to participate in this fruitful workshop which I 
benefited much from.  
The writing of the current manuscript is 
supported in part by funding from the Office of Naval Research.

\end{document}